\documentclass{webofc}\usepackage[varg]{txfonts} 
\usepackage{color}
\woctitle{QCD@Work 2018}\begin{document}\title{Strong Isospin
Breaking in Heavy-Meson Decay Constants: Employing Borelized QCD
Sum Rules in Local-Duality Limit}\author{Wolfgang Lucha\inst{1}
\fnsep\thanks{\email{Wolfgang.Lucha@oeaw.ac.at}}\and Dmitri
Melikhov\inst{1,2,3}\fnsep\thanks{\email{dmitri_melikhov@gmx.de}}
\and Silvano Simula\inst{4}\fnsep
\thanks{\email{simula@roma3.infn.it}}}\institute{Institute for
High Energy Physics, Austrian Academy of Sciences,
Nikolsdorfergasse 18,\\A-1050 Vienna, Austria\and D.~V.~Skobeltsyn
Institute of Nuclear Physics, M.~V.~Lomonosov Moscow State
University,\\119991 Moscow, Russia\and Faculty of Physics,
University of Vienna, Boltzmanngasse 5, A-1090 Vienna, Austria\and
INFN, Sezione di Roma Tre, Via della Vasca Navale 84, I-00146
Roma, Italy}

\abstract{Applying the QCD sum-rule machinery in the so-called
local-duality shape to heavy--light mesons reveals that, as a
consequence of the non-zero mass gap between up and down quarks,
the leptonic decay constants of the neutral and charged versions
of the $D,$ $D^\ast,$ $B$ and $B^\ast$ mesons differ by
approximately~$1\;\mbox{MeV}.$}\maketitle

\section{Target: Heavy-Pseudoscalar- and -Vector-Meson Decay
Constants}The discrepancy $(m_d-m_u)(2\;\mbox{GeV})\approx
2.5\;\mbox{MeV}$ \cite{PDG} of the masses $m_{u,d}$ of up and down
quarks is rather small compared to even the lightest meson masses
but causes a strong-isospin-breaking disparity of the
\emph{leptonic decay constants\/} of mesons comprising heavy quark
plus $u$ or $d$ quark.

We study pseudoscalar mesons $P_q$ and vector mesons $V_q,$
subsumed by $H_q,$ as heavy--light mesons formed by a heavy quark
$Q=c,b$ of mass $m_Q$ and a light quark $q=u,d,s$ of mass~$m_q,$
characterized by their masses $M_{H_q}=M_{P_q},M_{V_q},$
four-momenta $p,$ polarization~vectors $\varepsilon_\mu(p)$ in the
case of vector mesons $V_q,$ and leptonic decay constants
$f_{H_q}=f_{P_q},f_{V_q},$ defined according~to$$\langle0|\,\bar
q(0)\,\gamma_\mu\,\gamma_5\,Q(0)\,|P_q(p)\rangle={\rm i}\,f_{P_q}
\,p_\mu\ ,\qquad\langle0|\,\bar q(0)\,\gamma_\mu\,Q(0)\,
|V_q(p)\rangle=f_{V_q}\,M_{V_q}\,\varepsilon_\mu(p)\ .$$

The formalism of QCD sum rules \cite{QSR} provides an analytical
approach to systems governed by the strong interactions. For our
goals, it seems very promising to follow a path opened~by~a
limiting case allowing us to concentrate all nonperturbative
aspects within a single ingredient.

\section{Tool: (Borel-Transformed) QCD Sum Rule in its
Local-Duality Limit}
\begin{figure}[hbt]\centering\begin{tabular}{ccc}
\quad\includegraphics[scale=.40541]{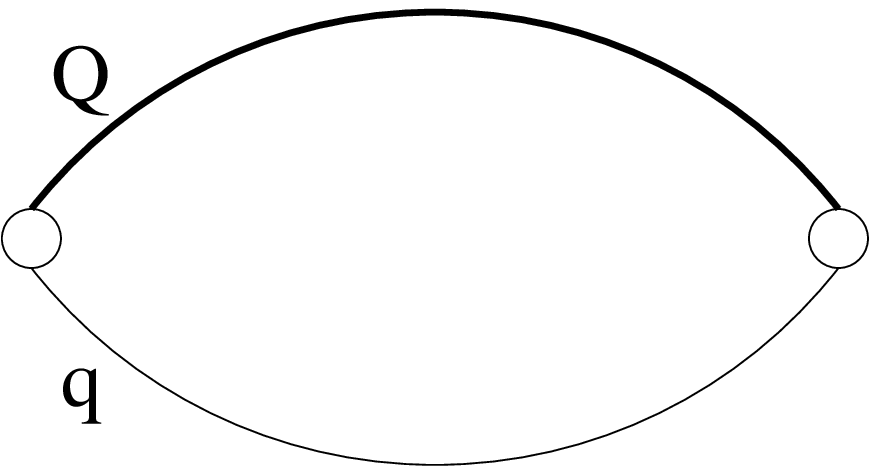}\quad&
\quad\includegraphics[scale=.40541]{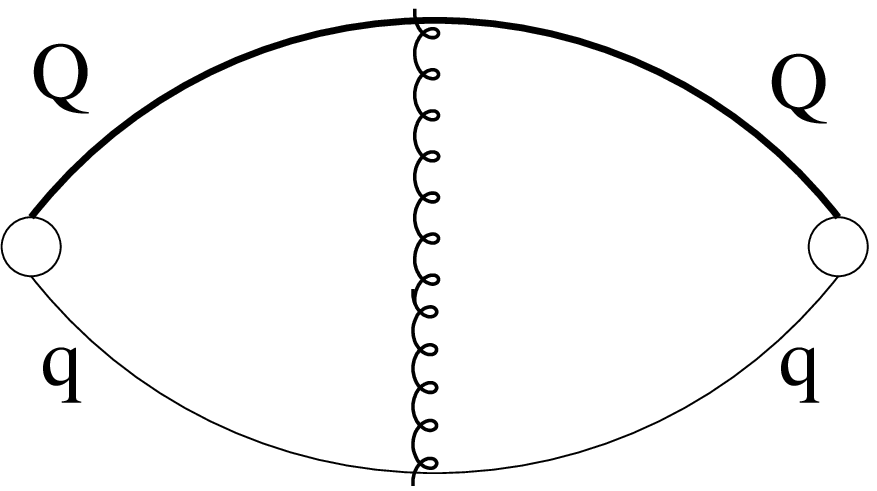}\quad&
\quad\includegraphics[scale=.40541]{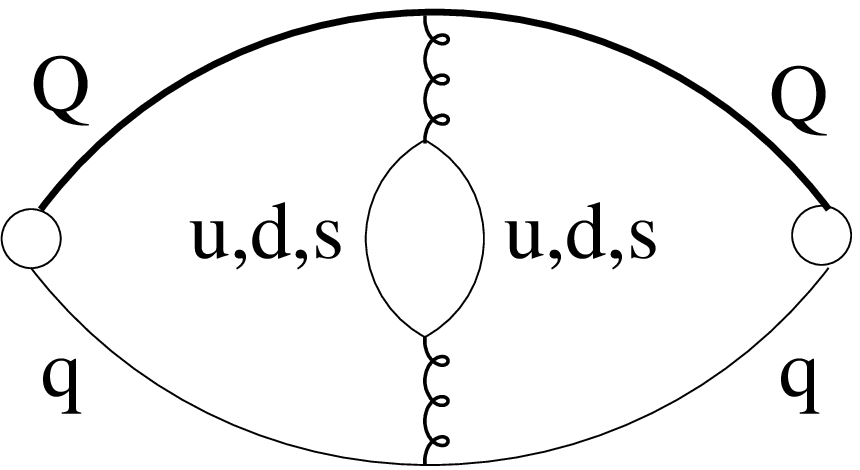}\quad
\\\quad(a)&\quad(b)&\quad(c)\end{tabular}\caption{Perturbative
expansion (\ref{SD}) of the spectral density
$\rho(s,m_Q,m_q,\alpha_{\rm s}\,|\,m_{\rm sea})$ in powers of the
strong coupling $\alpha_{\rm s}$: (a) $O(1),$ (b) $O(\alpha_{\rm
s}),$ (c) $O(\alpha_{\rm s}^2);$ sea quarks, and thus their
masses, start to contribute~at~$O(\alpha_{\rm s}^2).$}
\label{p}\end{figure}

A typical QCD sum rule \cite{QSR} extracted
from the two-point correlation function of appropriately chosen
interpolating local operators relates basic properties of the
hadron under consideration --- in our case, of any heavy--light
meson $H_q$ --- to the parameters of QCD. The latter are either of
fundamental nature, such as its strong fine-structure coupling
$\alpha_{\rm s}$ and all quark masses, or of effective type, such
as all the vacuum condensates $\langle\bar q\,q\rangle,\dots$
encoding nonperturbative aspects. For convenience, in order to
remove necessary subtraction terms and to suppress contributions
of hadronic excitations and continuum, usually a Borel
transformation from one's momentum variable to some Borel
parameter, called $\tau,$ is performed. The actual shape of such a
QCD sum rule depends on its detailed formulation, reflected by a
positive integer~exponent $N=0,1,\dots$:\begin{align}f_{H_q}^2
\left(M_{H_q}^2\right)^N\exp(-M_{H_q}^2\,\tau)&=\hspace{-2.4924ex}
\int\limits_{(m_Q+m_q)^2}^{s^{(N)}_{\rm
eff}(\tau,m_Q,m_q,\alpha_{\rm s})}\hspace{-3.09ex}{\rm d}s
\exp(-s\,\tau)\,s^N\,\rho(s,m_Q,m_q,\alpha_{\rm s}\,|\,m_{\rm
sea})\nonumber\\&+\Pi^{(N)}_{\rm power}(\tau,m_Q,m_q,\alpha_{\rm
s},\langle\bar q\,q\rangle,\dots)\ .\label{s}\end{align}On its QCD
side, the three important ingredients are the $\tau$-dependent
\cite{LMST1,LMST2,LMST3,LMST4,LMST5} \emph{effective threshold\/}
$s^{(N)}_{\rm eff}(\tau,m_Q,m_q,\alpha_{\rm s}),$ constructed such
as to guarantee mutual cancellation of perturbative-QCD and hadron
contributions above the threshold to the utmost achievable degree;
the perturbative \emph{spectral density\/}
$\rho(s,m_Q,m_q,\alpha_{\rm s}\,|\,m_{\rm sea}),$ experiencing
also the masses $m_{\rm sea}$ of the sea quarks and expressible as
an expansion (Fig.~\ref{p}) in powers of $\alpha_{\rm s}$
depending on the renormalization scale~$\mu,$\begin{align}
\rho(s,m_Q,m_q,\alpha_{\rm s}\,|\,m_{\rm sea})&=\rho_0(s,m_Q,m_q)
+\frac{\alpha_{\rm s}(\mu)}{\pi}\,\rho_1(s,m_Q,m_q,\mu)\nonumber\\
&+\frac{\alpha_{\rm s}^2(\mu)}{\pi^2}\,
\rho_2(s,m_Q,m_q,\mu\,|\,m_{\rm sea})+O(\alpha_{\rm s}^3)\
;\label{SD}\end{align}and the $\tau$-dependent \emph{power
corrections\/} $\Pi^{(N)}_{\rm power}(\tau,m_Q,m_q,\alpha_{\rm
s},\langle\bar q\,q\rangle,\dots)$ that subsume, together with the
effective threshold $s^{(N)}_{\rm eff}(\tau,m_Q,m_q,\alpha_{\rm
s}),$ all the nonperturbative manifestations of QCD.

The partitioning of nonperturbative contributions between power
corrections and effective threshold reflects, in general, the
details of the problem to which the QCD sum rule is applied. By
considering, however, the local-duality limit $\tau\to0$ of
Eq.~(\ref{s}) for the handpicked exponent $N=0$ such that all
power corrections vanish we may achieve to shift all of the
nonperturbative burden to the effective threshold. With all power
corrections gone, Eq.~(\ref{s}) simplifies to
\cite{LMSLD1,LMSLD2,LMSLD3}\begin{equation}f_{H_q}^2=
\hspace{-2.3ex}\int\limits_{(m_Q+m_q)^2}^{s_{\rm
eff}(m_Q,m_q,\alpha_{\rm s})}\hspace{-2.3ex}{\rm
d}s\,\rho(s,m_Q,m_q,\alpha_{\rm s}\,|\,m_{\rm sea})\equiv
\digamma(s_{\rm eff}(m_q),m_q\,|\,m_{\rm sea})\ .\label{f}
\end{equation}All spectral densities required as input are
known up to order $O(\alpha_{\rm s}\,m_q)$ and $O(\alpha_{\rm
s}^2\,m_q^0)$ \cite{d1,d2,d3,d4}. However, we first must ensure
the well-definiteness of the local-duality limit
$\tau\to0$~of~Eq.~(\ref{s}).

\begin{figure}[h]\centering
\includegraphics[scale=.67293,clip]{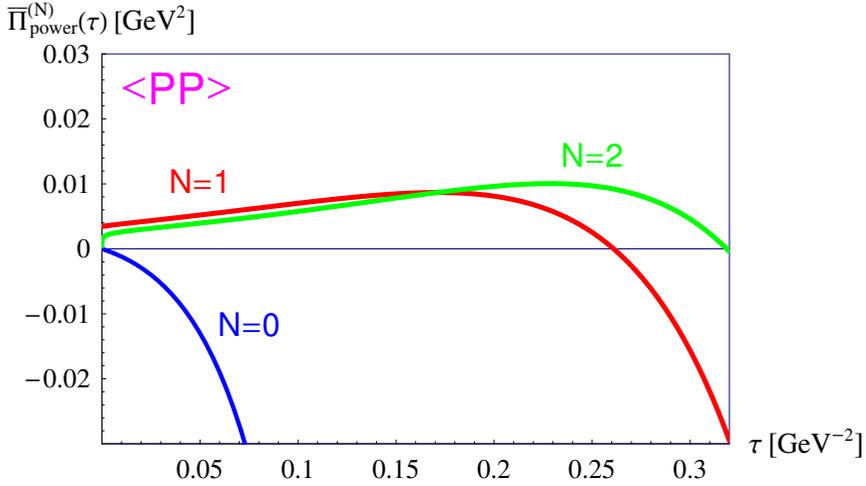}\\(a)\\[1ex]
\includegraphics[scale=.67293,clip]{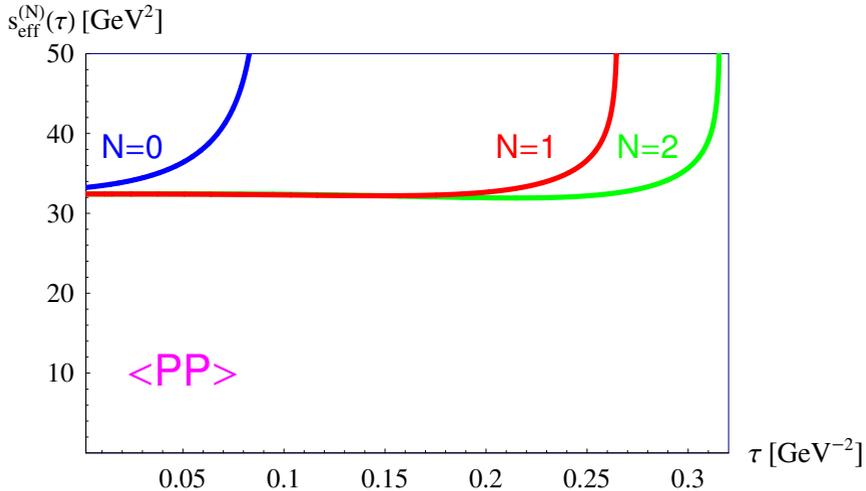}\\(b)
\caption{Dependence of (a) the power corrections $\Pi^{(N)}_{\rm
power}(\tau),$ rescaled according to Eq.~(\ref{pc}), as well as
(b) our effective threshold $s^{(N)}_{\rm eff}(\tau)$ in the QCD
sum rule for two pseudoscalar currents on the
Borel~variable~$\tau$.}\label{PP}\end{figure}

\begin{figure}[t]\centering
\includegraphics[scale=.67293,clip]{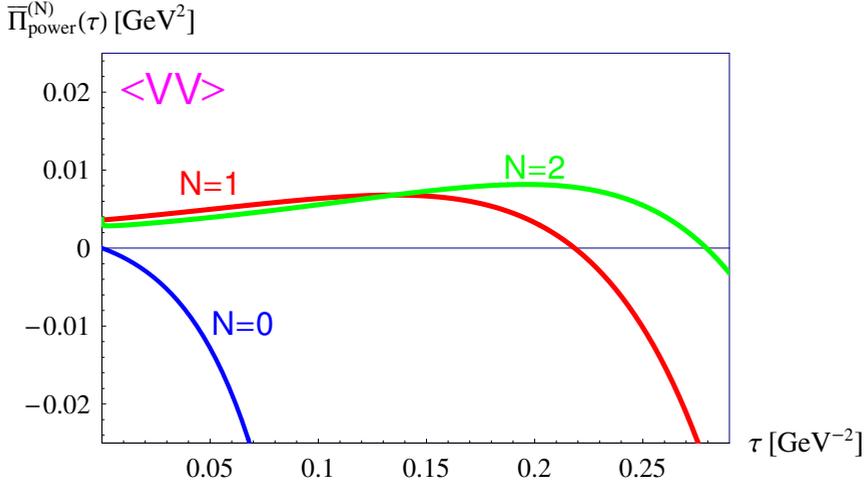}\\(a)\\[1ex]
\includegraphics[scale=.67293,clip]{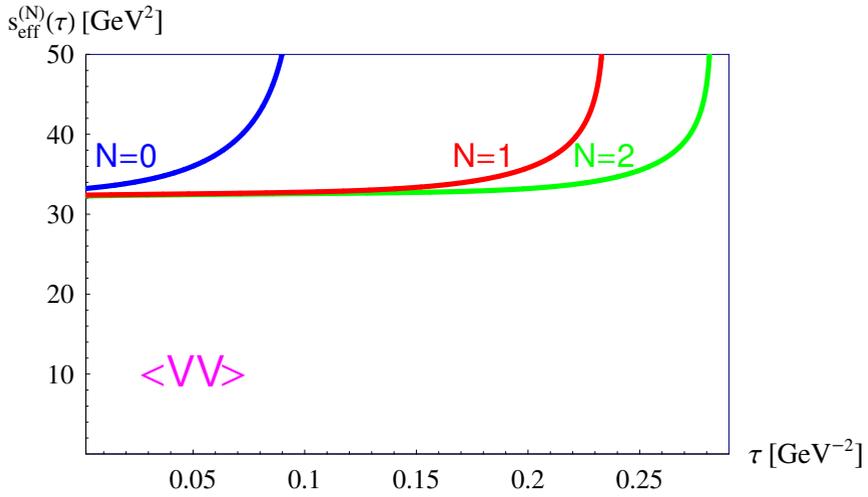}\\(b)
\caption{Dependence of (a) the power corrections $\Pi^{(N)}_{\rm
power}(\tau),$ suitably rescaled according to Eq.~(\ref{pc}), and
(b) our effective threshold $s^{(N)}_{\rm eff}(\tau)$ in the QCD
sum rule for two vector operators on the Borel parameter~$\tau$.}
\label{VV}\end{figure}

The dependence on the Borel parameter $\tau$ of our effective
threshold $s^{(N)}_{\rm eff}(\tau,m_Q,m_q,\alpha_{\rm s})$ and of
the power corrections $\Pi^{(N)}_{\rm
power}(\tau,m_Q,m_q,\alpha_{\rm s},\langle\bar q\,q\rangle,\dots)$
under the ``disguise'' of some~quantity\begin{equation}
\overline{\Pi}^{(N)}_{\rm power}(\tau)\equiv \Pi^{(N)}_{\rm
power}(\tau)\,\frac{\exp(M_{H_q}^2\,\tau)}{M_{H_q}^{2\,N}}\label{pc}
\end{equation}is explored, for the three most interesting values
$N=0,1,2$ of the exponent $N,$ in Fig.~\ref{PP}~for the
pseudoscalar mesons and in Fig.~\ref{VV} for the vector mesons.
For the cases $N=0$ and $N=1,$ the $\tau$ behaviour of the power
corrections at $\tau=0$ is regular: for $N=0,$ all
power~corrections~vanish,$$\lim_{\tau\to0}\Pi^{(N)}_{\rm
power}(\tau,m_Q,m_q,\alpha_{\rm s},\langle\bar q\,q\rangle,\dots)
=0\qquad\mbox{for}\quad N=0\ ;$$for $N=1,$ the power corrections
approach a finite value. In contrast to this, for the case~$N=2,$
due to an involvement of also $\log(\tau)$ terms the power
corrections exhibit singularities at $\tau=0.$ Hence, in our
pursuit of local-duality QCD sum rules the exponent of choice must
read $N=0.$

\section{Triumph: Local-Duality Outcomes for Decay-Constant
Differences}By collecting the available pieces of information
about the $m_q$ dependence of all the quantities entering in our
local-duality limit (\ref{f}) of the QCD sum rule (\ref{s}), we
extract the isospin-breaking difference $f_{H_d}-f_{H_u}$ from the
difference of an associated function $\digamma(s_{\rm
eff}(m_q),m_q\,|\,m_{\rm sea}),$ Eq.~(\ref{f}):\begin{align}&
\digamma(s_{\rm eff}(m_d),m_d\,|\,m_{\rm sea})-\digamma(s_{\rm
eff}(m_u),m_u\,|\,m_{\rm sea})\nonumber\\&=\digamma(s_{\rm
eff}(m_d),m_d\,|\,0)-\digamma(s_{\rm eff}(m_u),m_u\,|\,0)
+O(\alpha_{\rm s}^2\,(m_d-m_u))\ .\label{d}\end{align}

\begin{figure}[hbt]\centering
\includegraphics[scale=.48583,clip]{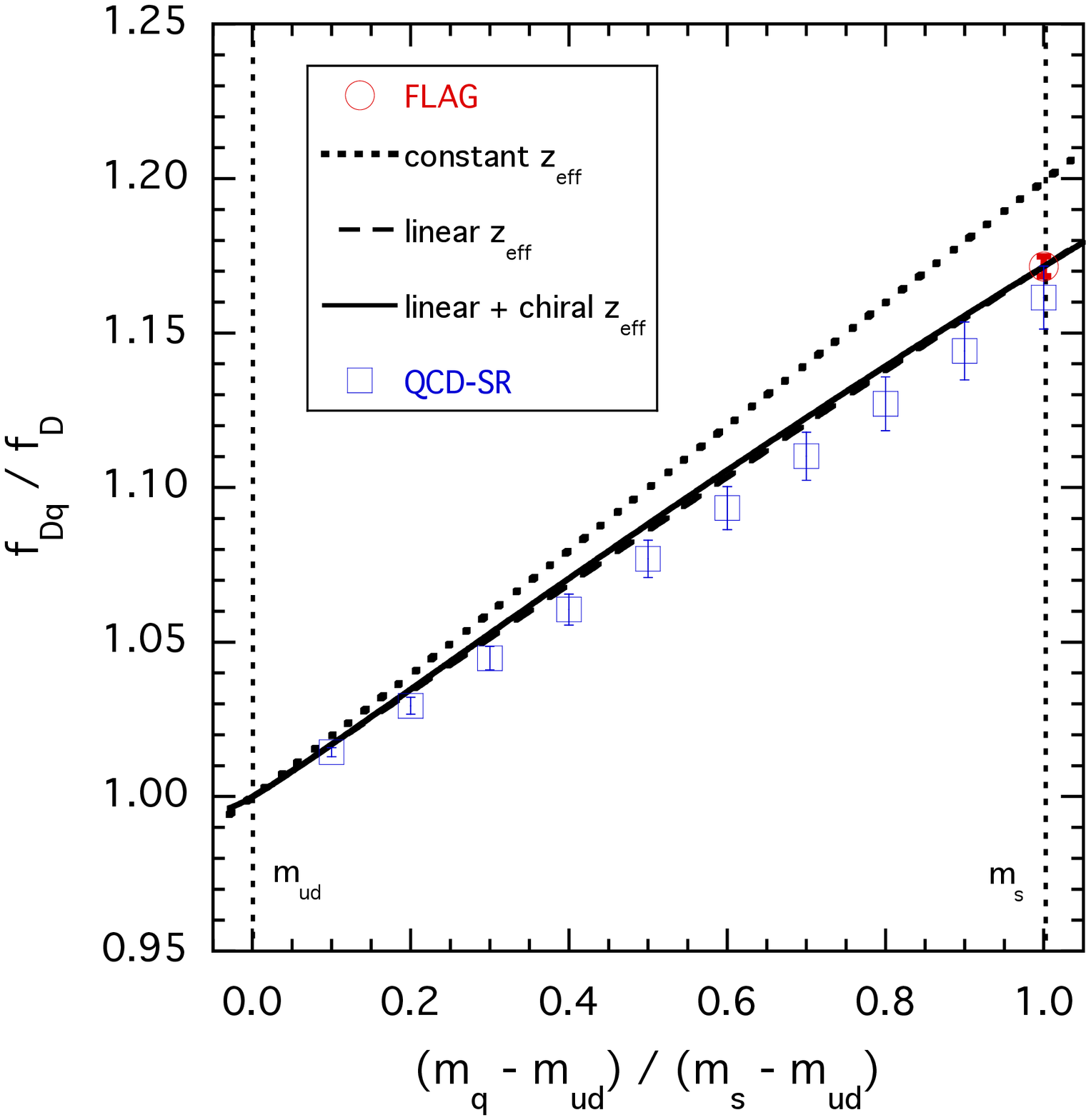}\\[2ex]
\includegraphics[scale=.48583,clip]{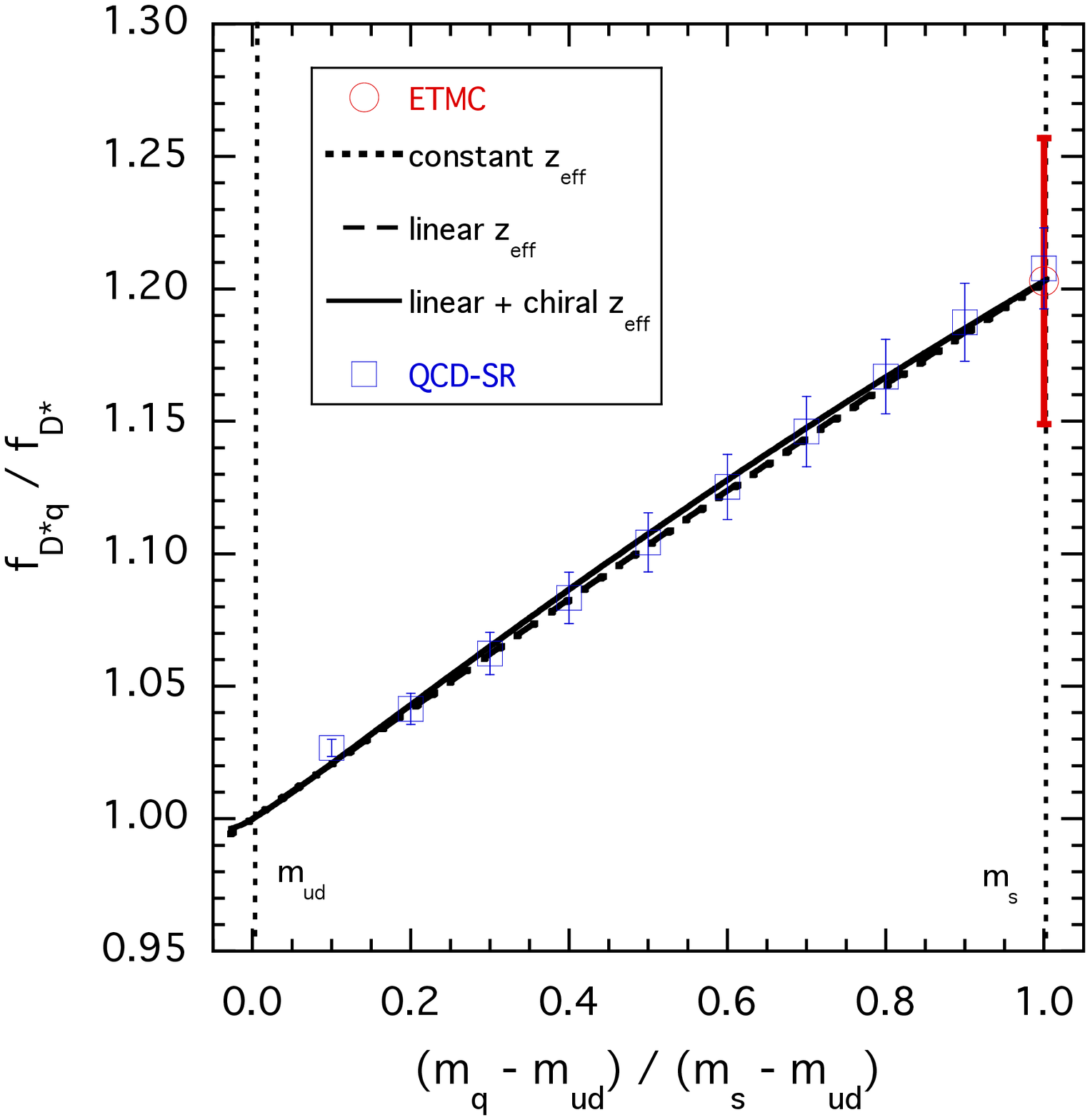}
\caption{Dependence of the heavy--light charmed-meson decay
constants, $f_{D^{(\ast)}_q}\equiv f_{D^{(\ast)}_q}(m_q),$
normalized to their values $f_{D^{(\ast)}}\equiv
f_{D^{(\ast)}_q}(m_{ud})$ at the average light-quark mass,
$m_{ud}\equiv(m_u+m_d)/2,$ on the (conveniently shifted and
rescaled) light-quark mass $(m_q-m_{ud})/(m_s-m_{ud}),$ as
inferred from three different assumptions about the behaviour of
our effective threshold $z_{\rm eff}\equiv\sqrt{s_{\rm
eff}}-m_c-m_q$ \cite{LMSLD1,LMSLD2,LMSLD3}, vs.\ previous findings
\cite{LMSIB}~($\textcolor{blue}{\square}$).}\label{D:f/f}\end{figure}
\begin{figure}[hbt]\centering
\includegraphics[scale=.48583,clip]{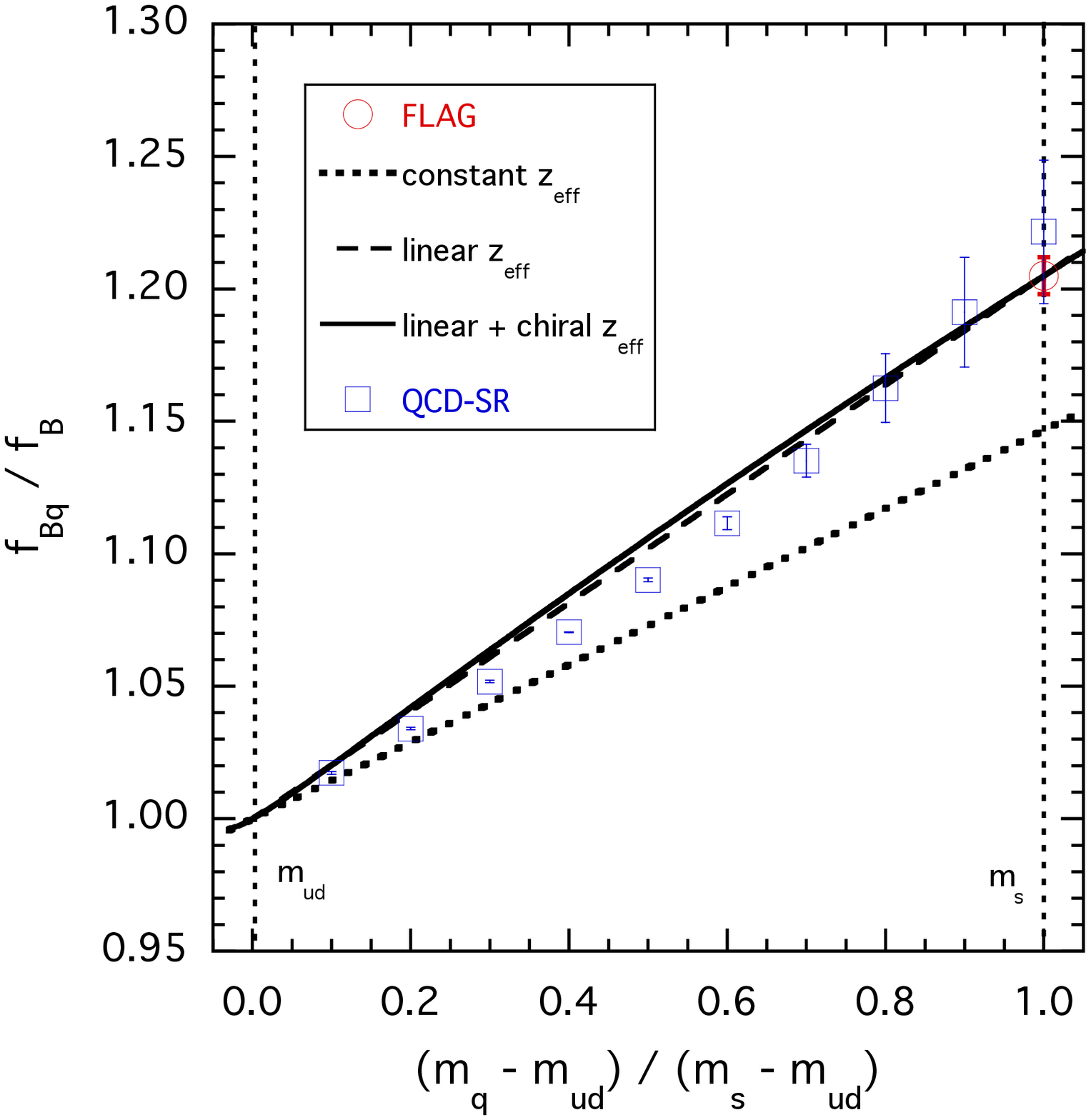}\\[2ex]
\includegraphics[scale=.48583,clip]{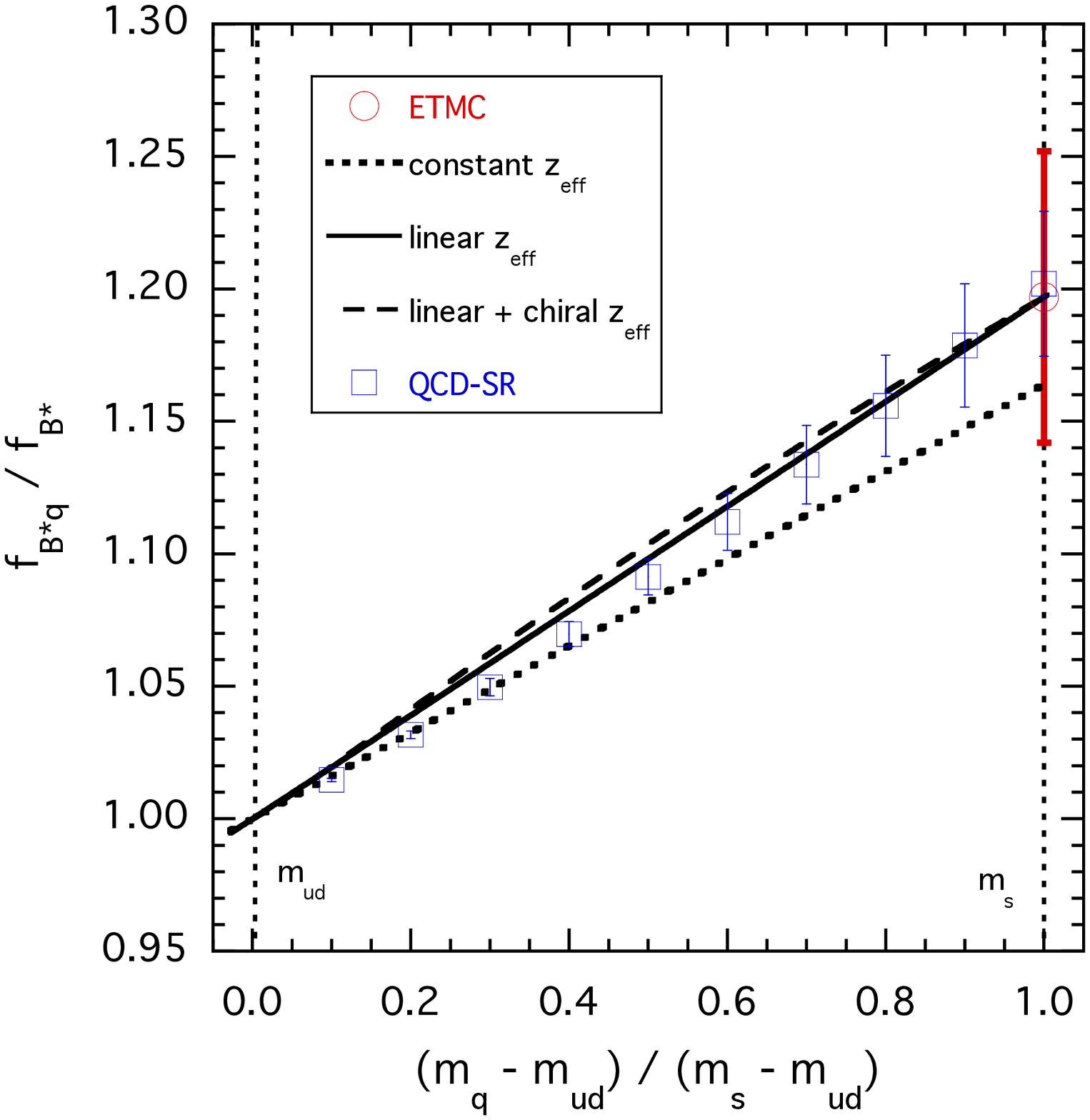}
\caption{Dependence of the heavy--light bottom-meson decay
constants $f_{B^{(\ast)}_q}\equiv f_{B^{(\ast)}_q}(m_q),$
normalized~to their values $f_{B^{(\ast)}}\equiv
f_{B^{(\ast)}_q}(m_{ud})$ at the average light-quark mass,
$m_{ud}\equiv(m_u+m_d)/2,$ on the --- still suitably shifted and
rescaled --- light-quark mass $(m_q-m_{ud})/(m_s-m_{ud}),$
inferred from three different assumptions about the behaviour of
our effective threshold $z_{\rm eff}\equiv\sqrt{s_{\rm
eff}}-m_b-m_q$ \cite{LMSLD1,LMSLD2,LMSLD3}, vs.\ previous findings
\cite{LMSIB}~($\textcolor{blue}{\square}$).}\label{B:f/f}\end{figure}
\clearpage

All dependence on the light-quark mass $m_q$ of our predictions
for leptonic decay constants $f_{H_q}$ by the local-duality QCD
sum rule (\ref{f}) originates in precisely two sources of
different~type:\begin{enumerate}\item The impact of the spectral
density can be trivially inferred from existing knowledge~due to
perturbation theory. Equation (\ref{d}) already takes into account
that in this difference~the contributions of \emph{strange\/} sea
quarks cancel. This reduces the uncertainties considerably.\item
The situation with the effective threshold is more delicate: we
model it by allowing~for~a continuous variation of $m_q$ within
$[0,m_s]$ and matching all results for $f_{H_q}$ to lattice~QCD in
three neighbouring ways, see Fig.~\ref{D:f/f} for charmed quarks
and Fig.~\ref{B:f/f} for bottom~quarks.\end{enumerate}In proper
naming, our results for the heavy--light meson decay-constant
differences are~\cite{LMSLD1,LMSLD2,LMSLD3}
\begin{align*}f_{D^\pm}-f_{D^0}&=(0.96\pm0.09)\;\mbox{MeV}\ ,&
f_{D^{*\pm}}-f_{D^{*0}}&=(1.18\pm0.35)\;\mbox{MeV}\ ,\\[.3277ex]
f_{B^0}-f_{B^\pm}&=(1.01\pm0.10)\;\mbox{MeV}\ ,&
f_{B^{*0}}-f_{B^{*\pm}}&=(0.89\pm0.30)\;\mbox{MeV}\ .\end{align*}
Strong isospin breaking in these heavy--light-meson decay
constants is pretty close to $1\;\mbox{MeV}.$

\section*{Acknowledgements}D.~M.\ is grateful for support by the
Austrian Science Fund (FWF) under project P29028-N27.

\end{document}